# Attention Masks Help Adversarial Attacks to Bypass Safety Detectors

*Yunfan Shi*

UCL

## ABSTRACT

Despite recent research advancements in adversarial attack methods, current approaches against XAI monitors are still discoverable and slower. In this paper, we present an adaptive framework for attention mask generation to enable stealthy, explainable and efficient PGD image classification adversarial attack under XAI monitors. Specifically, we utilize mutation XAI mixture and multitask self-supervised X-UNet for attention mask generation to guide PGD attack. Experiments on MNIST (MLP), CIFAR-10 (AlexNet) have shown that our system can outperform benchmark PGD, Sparsefool and SOTA SINIFGSM in balancing among stealth, efficiency and explainability which is crucial for effectively fooling SOTA defense protected classifiers.

*IndexTerms*— adversarial examples

## 1. INTRODUCTION

With the advancement of deep learning models for various tasks such as classification [1] and segmentation [2], they are increasingly deployed to numerous use cases [3] for better performance and autonomy. This brings to the issue of adversarial robustness [4] of all these autonomous systems with models learnt from large datasets that may be biased where some potential behaviours are difficult to discover and yet fatal to safety and reliability. Once malicious inputs are provided as input, the functionality can be problematic.

Recent advances in XAI [5] significantly help adversarial defense and monitoring which makes it necessary for novel adversarial attack that can fool XAI algorithms with similar outputs and related safety monitors which essentially compare both outputs via different mechanisms.

In practice, additional adversarial safety monitors [6] especially XAI based [15][17] is added to address the above issue without making large costly modifications to the already deployed model. For real time systems, attack efficiency is also crucial.

Our contributions are as follows:

1. Efficient explainable attention mask guided PGD: 17%faster and 0.01% less effective with 14% more stealth

2. Better XAI adversarial explanation without discriminator training using deconvolution layers under selfsupervision.

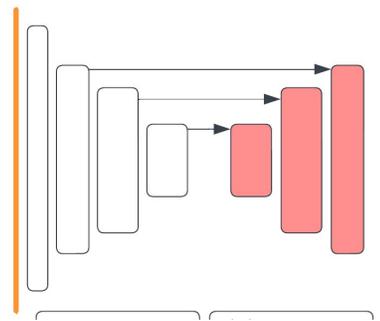

Fig. 1. Model architecture

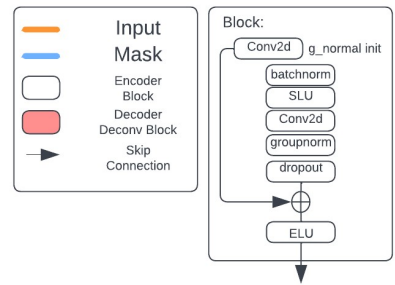

Fig. 2. Model architecture

### 1.1. Motivation

Current attack methods including PGD is still not stealthy enough to pass XAI monitors, not explainable enough and not efficient enough to fool target classifiers. We propose both XAI mixture mutation pgd attention mask generation algorithm without auxiliary model training and deep learning based X-UNet for attention mask generation to guide PGD for the discovery of query-efficient adversarial examples stealthy enough to pass XAI monitors and pixelwise saliency of such attacks under whitebox setting. We focus on attention mask generation via our X-U-Net with novel multi-task self-supervised loss function, activation function and weight initialization.

## 2. RELATED WORK

1. Adversarial monitor:

    Methods can be generally divided into reactive and proactive. reactive methods include adversarial detecting [7] and abstraction based [8] while proactive methods updates the model via retraining [9] or distillation [10]. These are reported to be empirical based defense, there are also Certified denfense with provable guarantee such as [11]. Monitoring can be applied on DNN inputs, intermediate values and outputs. Here, we focus on proactive XAI based monitor on DNN inputs given that it is reported by [12] [7] to be more effective when detecting current state of the art adversarial attacks.

2. Stealthy attack:

    Traditional efforts focus on reducing attack space usage, reducing epsilon, reducing step size which could not achieve a good balance between stealth and fooling rate. This task in itself is data dependent and is difficult to extract general pattern without injecting prior knowledge on certain datasets to refine the task [13]. To address this, current research utilizes generative deep learning such as GAN to produce sparse attack noise [14] and a hybrid approach which involves an algorithm which does not need an auxiliary model and training and thus can be more economical [15], together with GAN the system achieves better balance and flexibility in practice. Weights and masks are often applied for guidance using Hadamard Product [16]. In our setting, we fix space usage, epsilon and step size and combining ideas of sparsity, auxiliary model and hadamard product projection into an attention mask guided PGD attack.

3. Explainable attack:

    We focus on pixelwise saliency for attack explainability [17] as they can be of general use for attack guidance, explanation, defense and many other tasks. [18] mix Integrated Gradient with LRP via weighted relevance mask for better explainablity of transformers to outperform attention visualization. XAI mixture masks are used by [19] to improve the stealth of the attack as well as explainability. [20] propose attack towards explainable and efficient attack.

## 3. DESIGN

### 3.1. Metrics

Balance ratio for overall task effectiveness:

$$Avg(\lambda_1 * \delta stealth + \lambda_2 * \delta exp + \lambda_3 * (\frac{\sum mis}{t})) \quad (1)$$

Stealth: quantified by Cosine Similarity:

$$similarity = \frac{A * B}{max(||A||_2 * ||B||_2)} \quad (2)$$

Efficiency: quantified by Speed:

$$V = \frac{\sum mis}{t} \quad (3)$$

Explanation:

We conducted quantitative pixelwise saliency and empirical global pattern analysis.

### 3.2. Framework Design

#### 3.2.1. Workflow

First we use Integrated Gradient (IG) and LRP to get explanations of the target model on given dataset, and then we use mixture algorithm to generate partial attention masks which can be used to guide PGD attack on target model or as partial label to training X-UNet to further improve attack stealth and explainability.

#### 3.2.2. XAI safety monitor

The most common pattern is to first calculate the output value of adversarial examples via certain XAI methods [12] and then pass through a discriminator where different outputs result in a detection of adversary. Hence, the core task is to reduce the difference between clean image and adversarial examples across different metrics. Here, we simplify the task as reducing the Cosine Similarity [21] difference between adversarial examples and clean images.

#### 3.2.3. XAI mixture algorithm:

---

Algorithm 1 Mute (*lrp,ig,thresh*)

    for x in X do $a \leftarrow$ rand $b \leftarrow$ 1-a
      if a < thresh
        $a \leftarrow$ 1-a $b \leftarrow$ 1-b
      end if
      $mix \leftarrow 1 - (b * normalize(lrp) + a * ig)$
    end for
    return mix

---

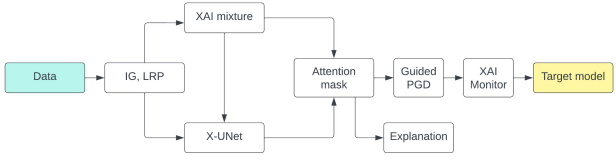

*Fig. 3. Data flow diagram*

### 3.2.4 X-UNet

We tried to modify every fundamental building blocks of the standard UNet as shown in Fig 1 & 2 and try to build a simple and elegant solution for our task. UNet configuration and novel building blocks inspired by [22] with state of the art components are dedicated for higher model discerning ability with efficiency in mind in the process of attention mask generation which ultimately makes our model output mask guided attack more stealthy. Loss:

$$\lambda_1 * L_1 loss(adv_m, data) + \lambda_2 * L_1 loss(mask, mix) + \lambda_3 * \delta acc \quad (4)$$

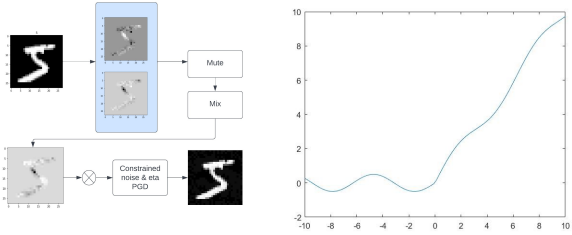

Fig. 4. Algorithm process

Fig. 5. Activation plot

This 3 term loss function is designed for multi-task selfsupervised learning where term 1 and 3 is the effect feedback regularization of term 2, together they form a close loop selfsupervision where the loss will drop (reward the learning) if the generated mask can cause the attack on validation data to be both stealthy (term 1 small) and effective (masked attack accuracy should be as low as vanilla attack).

### 3.2.5. Activation Function SLU(x, a=0.5):

$$max(0,x) + a * sin(x) \quad (5)$$

### 3.2.6. Convolution weight initialization

Normal distribution with variance:

$$Var = \frac{1}{\frac{inC+outC}{2} * KernelSize} \quad (6)$$

## 4. RESULTS & EVALUATION

### 4.1. SOTA Attack benchmark

The experiments here are designed to show our attention mask can improve vanilla PGD in terms of stealth, explainability and efficiency as well as our guided PGD can outperform SOTA attack methods.

- Attack efficiency benchmark:(average of 3 runs) 12%increase
  [clean accuracy baseline: 55%]
  Our method outperforms SOTA in terms of speed defined in metrics, despite SparseFool achieving best accuracy.

Table 1. Attack efficiency benchmark:

|             | Time | Accuracy | Loss |
|-------------|------|----------|------|
| Ours        | 14s  | 27.6%    | 2.4  |
| PGD         | 16s  | 23.6%    | 2.72 |
| SparseFool  | 161s | 7.03%    |      |
| SINIFGSM    | 51s  | 50%      |      |

- Attack stealth benchmark: (average of 3 runs) 10% increase
  Our method outperforms SOTA in terms of stealth, efficiency and explainability balance as in Table 4.1 though SINIFGSM provides 2.3% more stealth.

Table 2. Attack stealth benchmark:

|             | Stealth |
|-------------|---------|
| PGD         | 82%     |
| Ours        | 92%     |
| SparseFool  | 60.1%   |
| SINIFGSM    | 94.3%   |

- Attack explanation benchmark:
  Several key parts of digit 5 is assigned with higher weight and there are some background attack patterns discovered like the vertical line on the right hand side of the digit while attack some nearby pixels will not work well. As stated previously, the three layer background distribution transition and sparse alternating attention weight in the mask helps the attack to achieve a balance between stealth and efficiency. Note the 5 digit in the mask is a bit different in shape than clean image. By adding noise on digit shape edges, the attack can improve. The one at the background is discernible after zooming in but could not be discerned in original 28x28 resolution by human eye and can pass the XAI based safety monitor with 97% confidence.

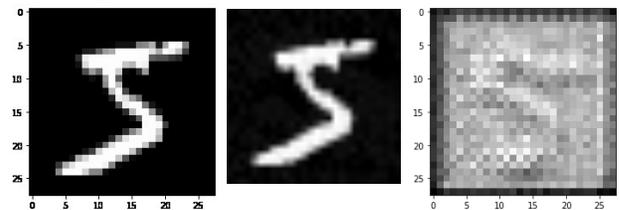

Fig. 6. MNIST sample clean    Fig. 7. MNIST sample adversarial    Fig. 8. MNIST sample attack mask

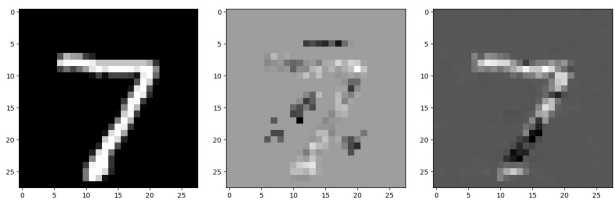

Fig. 9. MNIST Fig. 10. Sparse- Fig. 11. sample clean 2 fool IG SINIFGSM IG

Since SparseFool and SINIFGSM themselves are not explainable, we can still see uniform background with discrete not smooth distribution when passing through Integrated Gradient. Sparsity can be observed to some extent but as shown in the table above it is not as good as our solution. The same holds for stealth shown above in the table except for SINIFGSM with a 2% margin.

Table 3. Transfer on 3 metrics (best of 3 runs)

| (Per batch) | Acc | Time | Stealth |
|---|---|---|---|
| PGD | 0% | 0.07 | 87% |
| Ours | 0.01% | 0.06 | 98% |
| SparseFool | 11.7% | 676 | 89.2% |
| SINIFGSM | 11.7% | 5 | 65.71% |

When our algorithm without training is transferred onto AlexNet trained on CIFAR-10 dataset, results indicate that for stealth the guided PGD can still maintain a 11% gap with a 98% high confidence to fool monitor. The running time is 13% faster than vanilla PGD attack with same misclassification and very close accuracy drop so the efficiency is higher than MNIST. We observed a special effect on CIFAR-10 dataset where despite our PGD produces 30% less loss value but misclassification and accuracy is still good. Compared with MNIST, this could be because CIFAR-10 has larger image resolution 32x32 than MNIST 28x28 with RGB data on real life objects which are more complex than handwritten digit.

4.2. Limitation/future improvements

- Efficiency: The performance is still slow for first timeIG/LRP generation given data and model in whitebox setting. In this case, the time is not acceptable since the Integrated Gradient and LRP computation time is slower than attack itself by an order of 10. We can only assume an inference setting where data and model are fixed for a number of queries before recalculating XAI.

- Explanation: Our framework provides mostly qualitative explanation no enough mathematical guarantee compared to Integrated Gradient. A more robust metric should be employed.

5. CONCLUSION

In this paper, we presented our novel explainable efficient stealthy attack. Results have shown that our system can outperform benchmark vanilla PGD, Sparsefool and SINIFGSM in balance among stealth, efficiency and explainability. Results and pixelwise saliency provided by our novel framework can in turn benefit research for relevant tasks such as adversarial defense.

6. REFERENCES

[1] Alex Krizhevsky, Ilya Sutskever, and Geoffrey E Hinton, "ImageNet Classification with Deep Convolutional Neural Networks," in *Advances in Neural Information Processing Systems*. 2012, vol. 25, Curran Associates, Inc.

[2] Jonathan Long, Evan Shelhamer, and Trevor Darrell, "Fully Convolutional Networks for Semantic Segmentation," Mar. 2015, arXiv:1411.4038 [cs].

[3] Edward Korot, Zeyu Guan, Daniel Ferraz, Siegfried K. Wagner, Gongyu Zhang, Xiaoxuan Liu, Livia Faes, Nikolas Pontikos, Samuel G. Finlayson, Hagar Khalid, Gabriella Moraes, Konstantinos Balaskas, Alastair K. Denniston, and Pearse A. Keane, "Code-free deep learning for multi-modality medical image classification," *Nat Mach Intell*, vol. 3, no. 4, pp. 288–298, Apr. 2021, Number: 4 Publisher: Nature Publishing Group.

[4] Ian J. Goodfellow, Jonathon Shlens, and Christian Szegedy, "Explaining and Harnessing Adversarial Examples," Mar. 2015, arXiv:1412.6572 [cs, stat].

[5] Lukas Huber, Marc Alexander Kuhn, Edoardo Mosca, and Georg Groh, "Detecting word-level adversarial text attacks via SHapley additive exPlanations," in *Proceedings of the 7th Workshop on Representation Learning for NLP*, Dublin, Ireland, May 2022, pp. 156–166, Association for Computational Linguistics.

[6] Raul Sena Ferreira, Jean Arlat, Jeremie Guiochet, and Helene Waeselynck, "Benchmarking Safety Monitors for Image Classifiers with Machine Learning," in *2021 IEEE 26th Pacific Rim International Symposium on Dependable Computing (PRDC)*, Perth, Australia, Dec. 2021, pp. 7–16, IEEE.

[7] Erzhena Tcydenova, Tae Woo Kim, Changhoon Lee, and Jong Hyuk Park, "Detection of Adversarial Attacks in AI-Based Intrusion Detection Systems Using Explainable AI," *Human-centric Computing and Information Sciences*, vol. 11, no. 0, pp. 1–1, Sept. 2021.

[8] Thomas A. Henzinger, Anna Lukina, and Christian Schilling, "Outside the Box: AbstractionBased Monitoring of Neural Networks," Feb. 2020, arXiv:1911.09032 [cs, stat].

[9] Yang Bai, Yuyuan Zeng, Yong Jiang, Yisen Wang, ShuTao Xia, and Weiwei Guo, "Improving Query Efficiency of Black-Box Adversarial Attack," in *Computer Vision – ECCV 2020*, Andrea Vedaldi, Horst Bischof, Thomas Brox, and Jan-Michael Frahm, Eds., Cham, 2020, Lecture Notes in Computer Science, pp. 101–116, Springer International Publishing.

[10] Gongfan Fang, Jie Song, Chengchao Shen, Xinchao Wang, Da Chen, and Mingli Song, "Data-Free Adversarial Distillation," Mar. 2020, arXiv:1912.11006 [cs, stat].

[11] Trung Le, Anh Tuan Bui, Le Minh Tri Tue, He Zhao, Paul Montague, Quan Tran, and Dinh Phung, "On Global-view Based Defense via Adversarial Attack and Defense Risk Guaranteed Bounds," in *Proceedings of The 25th International Conference on Artificial Intelligence and Statistics*. May 2022, pp. 11438–11460, PMLR, ISSN: 2640-3498.

[12] Gil Fidel, Ron Bitton, and Asaf Shabtai, "When explainability meets adversarial learning: Detecting adversarial examples using shap signatures," 2019.

[13] Zhenhua Chen, Chuhua Wang, and David Crandall, "Semantically Stealthy Adversarial Attacks Against Segmentation Models," 2022, pp. 4080–4089.

[14] Run Wang, Ziheng Huang, Zhikai Chen, Li Liu, Jing Chen, and Lina Wang, "Anti-Forgery: Towards a Stealthy and Robust DeepFake Disruption Attack via Adversarial Perceptual-aware Perturbations," June 2022, arXiv:2206.00477 [cs].

[15] Xingyu Zhou, Yi Li, Carlos A. Barreto, Jiani Li, Peter Volgyesi, Himanshu Neema, and Xenofon Koutsoukos, "Evaluating Resilience of Grid Load Predictions under Stealthy Adversarial Attacks," in *2019 Resilience Week (RWS)*, Nov. 2019, vol. 1, pp. 206–212.

[16] Jin-Hwa Kim, Kyoung-Woon On, Woosang Lim, Jeonghee Kim, Jung-Woo Ha, and Byoung-Tak Zhang, "Hadamard product for low-rank bilinear pooling," *arXiv preprint arXiv:1610.04325*, 2016.

[17] P Charantej Reddy, Aditya Siripuram, and Sumohana S. Channappayya, "Can perceptual guidance lead to semantically explainable adversarial perturbations?," 2021.

[18] Hila Chefer, Shir Gur, and Lior Wolf, "Transformer Interpretability Beyond Attention Visualization," 2021, pp. 782–791.

[19] Maximilian Noppel, Lukas Peter, and Christian Wressnegger, "Backdooring Explainable Machine Learning," Apr. 2022.

[20] Yucheng Shi, Yahong Han, Quanxin Zhang, and Xiaohui Kuang, "Adaptive iterative attack towards explainable adversarial robustness," *Pattern Recognition*, vol. 105, pp. 107309, Sept. 2020.

[21] Fan Wang and Adams Wai Kin Kong, "Exploiting the Relationship Between Kendall's Rank Correlation and Cosine Similarity for Attribution Protection," *Advances in Neural Information Processing Systems*, vol. 35, pp. 20580–20591, Dec. 2022.

[22] Karen Simonyan and Andrew Zisserman, "Very deep convolutional networks for large-scale image recognition," 2015.